\documentclass[twocolumn]{aastex62}

\graphicspath{{./}{figures/}}

\submitjournal{ApJ}

\shorttitle{QED-reconnection}
\shortauthors{Schoeffler et al.}

\begin{document}

\title{Bright gamma-ray flares powered by magnetic reconnection in QED-strength magnetic fields}

\correspondingauthor{K. M. Schoeffler}
\email{kevin.schoeffler@tecnico.ulisboa.pt}

\author{K. M. Schoeffler}
\author{T. Grismayer}
\affil{GoLP/Instituto de Plasmas e Fus\~ao Nuclear,
Instituto Superior T\'ecnico,\\
Universidade de Lisboa, 1049-001 Lisboa, Portugal\\}
\author{D. Uzdensky}
\affiliation{Center for Integrated Plasma Studies, Physics Department,\\ 
University of Colorado, Boulder CO 80309, USA}
\affiliation{Institute for Advanced Study, Princeton, NJ 08540, USA}
\author{R. A. Fonseca}
\affiliation{DCTI/ISCTE Instituto Universit\'ario de Lisboa, 1649-026 Lisboa, Portugal}
\affiliation{GoLP/Instituto de Plasmas e Fus\~ao Nuclear,
Instituto Superior T\'ecnico,\\
Universidade de Lisboa, 1049-001 Lisboa, Portugal\\}
\author{L. O. Silva}
\affil{GoLP/Instituto de Plasmas e Fus\~ao Nuclear,
Instituto Superior T\'ecnico,\\
Universidade de Lisboa, 1049-001 Lisboa, Portugal\\}



\begin{abstract}

Strong magnetic fields in magnetospheres of neutron stars (especially magnetars) and other astrophysical objects may release their energy in violent, intense episodes of magnetic reconnection. While reconnection has been studied extensively, the extreme field strength near neutron stars introduces new effects: radiation cooling and electron-positron pair production. Using massively parallel particle-in-cell simulations that self-consistently incorporate these new radiation and quantum-electrodynamic effects, we investigate relativistic magnetic reconnection in the strong-field regime.  
We show that reconnection in this regime can efficiently convert magnetic energy to X-ray and gamma-ray radiation and thus power bright high-energy astrophysical flares. Rapid radiative cooling causes strong plasma and magnetic field compression in compact plasmoids. In the most extreme cases, the field can approach the quantum limit, leading to copious pair production. 

\end{abstract}



\section{Introduction} \label{sec:intro}

Magnetic reconnection, a sudden, violent rearrangement of magnetic field
leading to a rapid release of magnetic energy, powers many spectacular flaring
events in space and astrophysical plasmas, e.g., solar flares, geomagnetic
storms, and high-energy flares from various astrophysical objects~\citep{Zweibel}.  
In the most extreme sources, such as magnetar and pulsar magnetospheres and gamma-ray bursts (GRBs), the reconnecting magnetic field is so strong that its dissipation leads to powerful $\gamma$-ray emission and copious $e^-e^+$ pair production. 
Both of these effects can, in turn, significantly affect the reconnection process itself as well as its observational appearance~\citep{Uzdensky-2011, Uzdensky-2016, Beloborodov-2017}.  
Until now, however, these radiation and quantum electrodynamic (QED) processes have not yet been fully considered in a first-principles calculation. Numerical studies of reconnection utilizing particle-in-cell (PIC) simulations have only recently started to incorporate synchrotron cooling ~\citep{Hoshino,Cerutti2013,Cerutti2014,Nalewajko_etal-2015}. Here we report the results of the first systematic {\it ab initio} study of relativistic magnetic reconnection which self-consistently includes 
nonlinear Compton radiation (which reduces to synchrotron emission when the magnetic field greatly exceeds the electric field), as well as pair production by the decay of MeV gamma-photons propagating across strong magnetic and electric fields. 


The main quantity that governs the relative strength of QED effects and radiation cooling is the magnetic field strength~$B_0$. QED effects can be conveniently characterized by the relativistic invariant $\chi_e \approx p_\perp B_0/(m_e c B_Q)$~\citep{Klepikov, Erber, Ritus_thesis}, where $B_Q \equiv m_e^2 c^3/e\hbar \simeq 4.4 \times 10^{13}\, {\rm G}$
$=E_Q$ (in Gaussian units) is the QED (Schwinger) field. This dimensionless quantity corresponds to the electric field $E$ in the rest frame of an electron (or positron) with momentum $p_\perp$ perpendicular to~$B_0$, normalized to~$E_Q$, and can be generalized for a photon with the same momentum.
The rate of radiation cooling can be expressed as $\dot{\gamma}_{\rm rad} /\gamma \sim \alpha_{\rm fs} \chi_e {\Omega_c}$ (where $\alpha_{\rm fs} \equiv e^2/\hbar c$
is the fine structure constant and $\Omega_c\equiv eB_0/m_ec$ is the classical cyclotron frequency). Significant cooling thus occurs when the cooling time is comparable to a characteristic time of the system $t$ (e.g. the light crossing time).
Our study involves a broad range of reconnecting magnetic fields~$B_0$ 
spanning across three distinct physical regimes:

(1)  {\it Classical non-radiative relativistic reconnection} occurs in relatively weak magnetic fields so that the local average $\chi_e$-parameter is very small everywhere, $\left<\chi_e\right> <\left(\Omega_c t\right)^{-1}$, and hence neither radiative cooling nor QED pair creation is important. 
This regime is relevant to PWN and winds and magnetospheres of weak pulsars~\citep{Lyubarsky,KirkHarris} and it serves as the baseline for our comparative study of radiative and QED effects on reconnection.

(2) {\it Radiative relativistic reconnection} occurs in moderately strong magnetic fields, $\left(\Omega_c t\right)^{-1} < \left<\chi_e\right> \ll
1$, where strong radiative cooling significantly affects the overall energetics and dynamics of reconnection but pair production remains insignificant~\citep{Hoshino, Cerutti2013, Cerutti2014, Uzdensky_Spitkovsky-2014, Uzdensky-2016}. 
This regime is applicable to the equatorial current sheet beyond the light cylinder in magnetospheres of bright gamma-ray pulsars like the Crab~\citep{Lyubarsky-1996, Uzdensky_Spitkovsky-2014,  Uzdensky-2016, Cerutti_etal-2016,Philippov_Spitkovsky-2018}.

(3) {\it The QED regime of radiative relativistic reconnection with pair creation} occurs in strong magnetic fields approaching the quantum (Schwinger) field, e.g., $\left<\chi_e\right> \sim 1$; this field is so strong that the mean free path of the produced gamma-ray photons with respect to QED one-photon pair production becomes short and large numbers of pairs can be produced. This regime is applicable to the most extreme astrophysical objects: GRB jets and magnetars~\citep{Thompson, Parfrey2013, Uzdensky-2011, McKinney_Uzdensky-2012, Uzdensky-2016, Kaspi2017}; to a lesser extent, it is also relevant  to the most powerful gamma-ray pulsars including the Crab~\citep{Lyubarsky-1996,Philippov_Spitkovsky-2018}. 

In this study, we show that there are significant qualitative differences between reconnection in the classical, radiative, and QED regimes.

\section{Simulation setup} \label{sec:setup}

Motivated by these considerations, we conducted a two-dimensional (2D) PIC study of relativistic  reconnection in a pair plasma, taking advantage of the OSIRIS framework~\citep{OSIRIS}. OSIRIS self-consistently includes radiation and the QED process of pair production by a single $\gamma$-ray photon  
propagating across a strong electromagnetic field~\citep{GrismayerPOP,GrismayerPRE}.
In this section, we present only the basic description of our simulation setup, while a more detailed discussion can be found in Appendix~A.

We simulate a 2D double relativistic Harris~\citep{Harris1962, KirkHarris} initial equilibrium with periodic boundary conditions.
The computational domain is initially filled with a relativistically hot background electron-positron plasma with uniform density (of each species) $n_b$ and temperature $T_b =4 m_e c^2$, chosen to yield a high upstream plasma magnetization $\sigma_h \equiv B_0^2/4 \pi (2n_b) h_b = 6.44$, where $B_0$ is the reconnecting field and $h_b$ is the relativistic enthalpy per particle ($h_b\approx 4T_b$ for ultrarelativistic temperatures).
We also include a small out-of-plane ($\hat{z}$) uniform guide magnetic field $B_G = 0.05 B_0$. 
In addition to the uniform background, we introduce two anti-parallel initial Harris current layers, each lying in a $y={\rm const}$ plane and carrying electric current in the $\pm \hat{z}$ direction. The layers have central electron and positron densities $n_0 = 10 n_b$, temperature $T_0=6.92 m_e c^2$, and a half-thickness $\delta = 2.55 \rho_L$. Here, our main fiducial length scale $\rho_L \equiv \gamma_T m_e c^2/eB_0 = \gamma_T c/\Omega_c$ is defined as the Larmor radius of a background particle with a Lorentz factor corresponding to the peak of the initial upstream relativistic Maxwell-J\"uttner distribution,  $\gamma_T \equiv 2T_b/m_e c^2$. 

A novel feature of our simulations is the self-consistent inclusion of radiation emission as well as propagation and pair-production absorption of the radiated photons. Our treatment of radiation emission has two alternative implementations, employed depending on the emitting particle's energy.
For low-energy particles below a certain energy threshold ($\gamma < 10$) we use a continuous description, with the radiation back-reaction accounted for classically using the Landau-Lifshitz model~\citep{LandauLifshitz} for radiative drag force, while we keep track of the total radiated energy. 
For more energetic particles, however, we model the emission as nonlinear Compton scattering in strong electromagnetic fields, accounting for the production of discrete hard photons (with resulting photon energies above~keV). The radiation recoil on the emitting particles is self-consistently implemented via momentum conservation between the hard photon and the particle. We treat these hard photons as computational particles that are propagated in the simulation ballistically in straight lines at the speed of light. At each time step, each of the $>$ MeV photons has a certain probability rate (depending on the $\chi_\gamma$ of the photon) to be converted into an $e^+e^-$ pair (see Appendix~C.1); when this happens, the photon is removed and a new electron-positron pair is deposited into the simulation, satisfying momentum conservation. 

Our typical simulation domain size is $2 L_x \times 2 L_y = 943\rho_L\times 943\rho_L$, with $3840 \times 3840$ computational cells of size $\Delta x = \Delta y = 0.246\rho_L$, initially with $16$ particles per species in each cell, with a total of nearly $0.5 \times 10^9$ particles.  In our strong-radiation runs, however, large numbers of photons (as well as some secondary pairs) are created in the course of the simulation, so that the total number of simulation particles (including photons) grows and reaches up to $3\times10^{9}$. The simulations are typically run for about 4 light crossing times~$L_y/c$ ($15088~\Omega_c^{-1}$), with a time step of 
$\Delta t = 0.58 \Delta x/c =\ 0.14 \rho_L/c = 0.14 \gamma_T \Omega_c^{-1}$.

We conducted a series of simulations with magnetic field strengths spanning the range $B_0/B_Q = 4.53 \times 10^{-6} - 4.53 \times 10^{-3}$. 
For clarity, however, we present only three representative cases here, each illustrating one of the above-described distinct physical regimes:
(1) the classical case $B_0/B_Q = 4.53 \times 10^{-6}$ where $\left<\chi_e\right> < \left(\Omega_c t\right)^{-1}$; 
(2) the radiative case $B_0/B_Q = 4.53 \times 10^{-4}$ where $\left(\Omega_c t\right)^{-1} \ll \left<\chi_e\right> \ll 1$; 
and (3) the QED case $B_0/B_Q = 4.53 \times 10^{-3}$ where $\left<\chi_e\right>$ can reach $\sim 1$ over the course of the reconnection process.

\section{Results} \label{sec:results}

In all our simulations, the reconnection process develops along a familiar sequence of events. 
First, each of the initial two current layers becomes unstable to the tearing instability that quickly breaks it up into a chain of magnetic islands (plasmoids) separated by small secondary current sheets containing reconnecting X-points. Next, as the islands grow and become nonlinear, they start moving along the layer and merge with each other in a hierarchical fashion, until eventually only one big island is left in each layer.  While this general morphological evolution is the same, there are substantial differences in dynamics, energetics, and radiative appearance between the three cases. 

\begin{figure*}[ht!]
\plotone{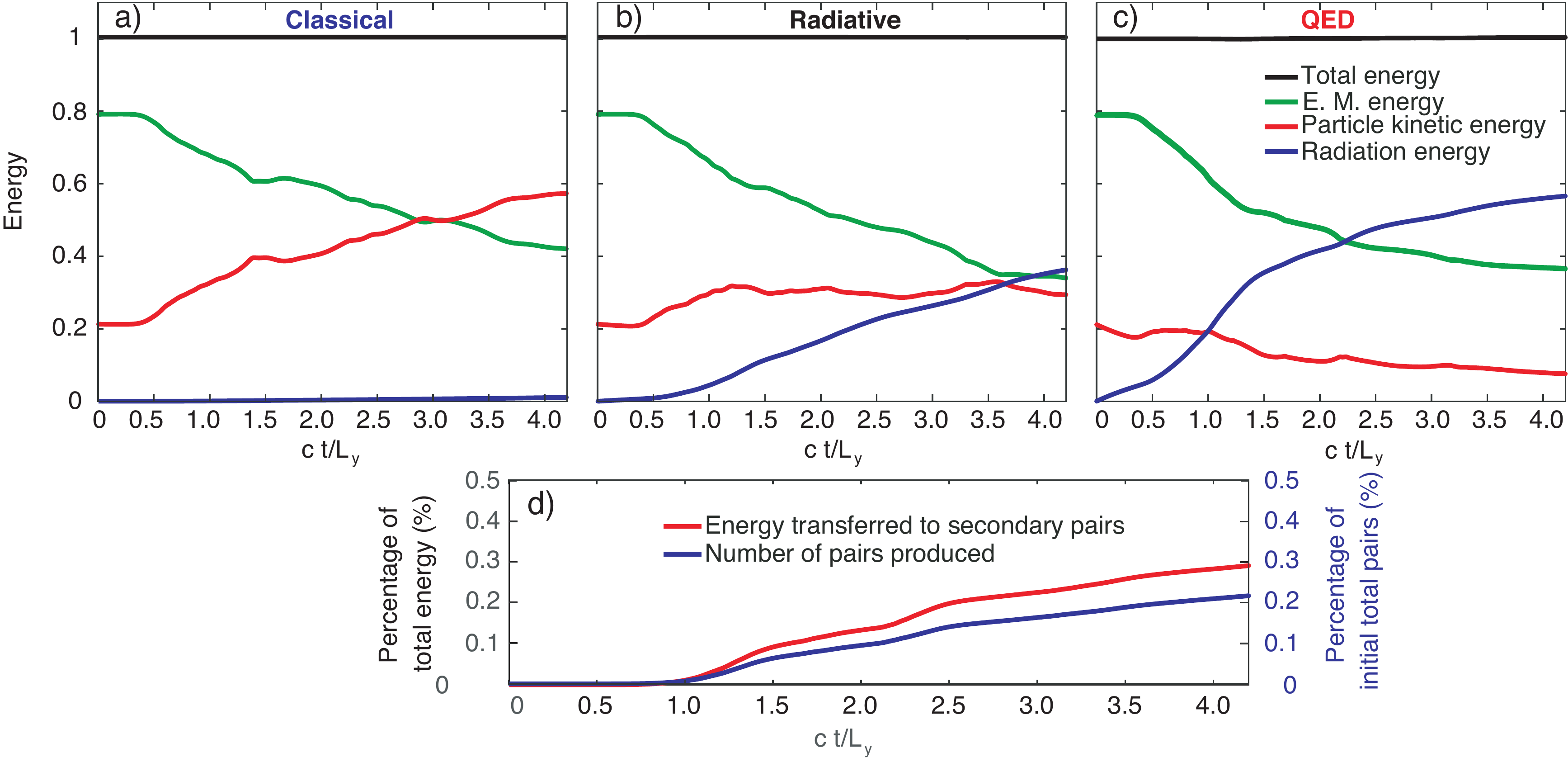}
\caption{Top raw (panels a-c): 
time evolution of various key energy components integrated over the system's volume, for the classical case (a), the radiative case (b), and the QED case~(c). 
The electric and magnetic field energy is shown in green, the kinetic energy of the electrons
and the positrons (including newly created pairs) in red, and the energy emitted as radiation is blue. 
The total energy of the system is shown in black. 
Bottom panel (d) shows the percentage of the energy that went into pair production (in red) and the relative number fraction of the produced pairs (in blue), for the QED case.}
\end{figure*}

The first manifestation of the differences between these cases can be seen in the time evolution of the system's energy content (see Fig.~1). Reconnection converts free energy of a reversing magnetic field into the kinetic energy of the particles (which can take the form of bulk flows, plasma heating, and nonthermal particle acceleration), as is clearly seen in the classical case shown in Fig.~1a. In the radiative case, however, shown in Fig.~1b, the energized particles quickly and efficiently radiate their energy and so most of the released magnetic energy is promptly transferred to hard photons, while the particle kinetic energy saturates at a relatively low constant level. This radiative cooling effect is also present, and is even stronger, in the QED case  (Fig.~1c).  In addition, however, a small but noticeable portion ($\sim 0.3\%$) of the total energy powers secondary pair production in the QED case, increasing the total number of electrons and positrons in the domain by a similar percentage (see Fig.~1d). 


Importantly, while the energy going to the secondary pair production is overall small even in the QED case, this energy conversion channel is highly concentrated in the cores of magnetic islands (plasmoids), which comprise only about $0.5\%$ of the total area of the simulation. Thus, pair production accounts for a significant ($\sim 1$) fraction of the local energy budget there. 
Likewise, the number density of newly produced pairs inside the plasmoids becomes comparable to~$n_b$.

\begin{figure*}[ht!]
\plotone{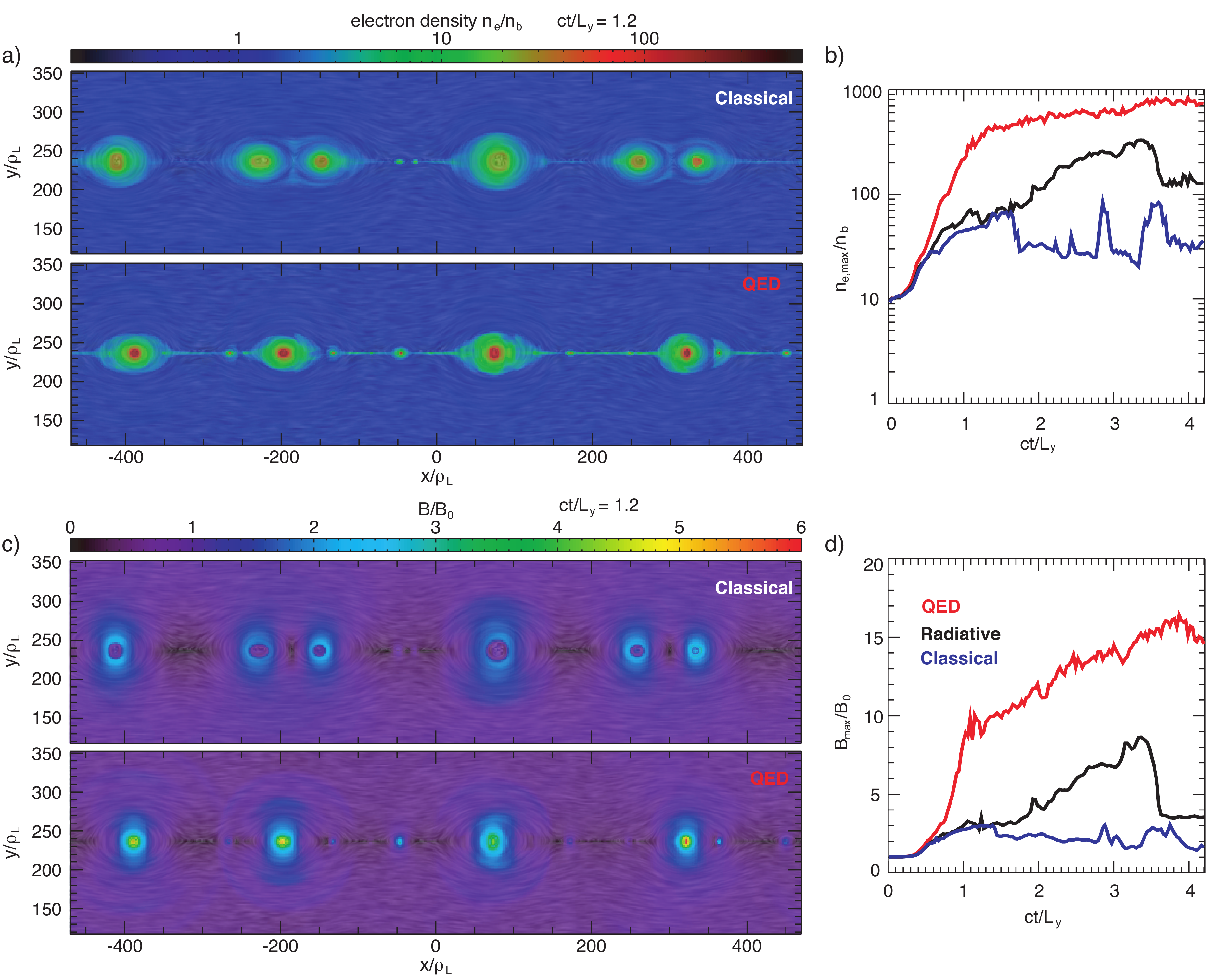}
\caption{Panel (a) shows the maps of the electron density $n$ at $t = 1.2 L_y/c$ for the classical (top sub-panel) and QED (bottom sub-panel) cases around the current sheet on the upper half of the simulation domain; the gray-scale texture overlay represents the in-plane ($xy$) magnetic field lines.
The evolution of the maximum electron density is shown in panel (b) for the QED case in red, the radiative case in black, and the classical case in blue. 
Panels (c) and (d) are the same as (a) and (b), respectively, but for the total magnetic field strength~$|\bf{B}|$.}
\end{figure*}

This strong concentration of pair production in the plasmoids is just one manifestation of an important and nontrivial general feature of reconnection: generation of strong inhomogeneities in the magnetic field and, especially, in the plasma density and pressure. These inhomogeneities have important consequences for all three regimes. They can be seen in Fig.~2 (a and c) which shows density and magnetic field maps in the classical and QED runs at $t = 1.2 L_y/c$, after the tearing instability has reached the non-linear stage.  

In all cases, the magnetic islands are filled with plasma and reconnected magnetic flux, leading to concentrated density and magnetic field [c.f. \citet{Sironi2006}].
A pinch equilibrium~\citep{Bennett} is established inside each island, with the inward magnetic tension balanced by the enhanced central plasma pressure (and also the pressure of the compressed guide field).  However, in the radiative and QED regimes, the high energies of accelerated particles, in conjunction with the strongly compressed magnetic fields, lead to powerful radiative cooling causing the pressure balance in the plasmoids to evolve towards even stronger compression.  This, in turn, results in an even stronger magnetic field amplification, further enhancing radiative cooling and thus leading to a positive feedback loop. 

As shown in Fig.~2 (b and d), the peak density and magnetic field enhancements reached in the radiative case ($n/n_b=300$, $B/B_0 = 8$) and, especially, in the QED case ($n/n_b=900$, $B/B_0 = 16$, i.e., reaching about $7\%$ of~$B_Q$) are significantly stronger than those found in the classical case ($n/n_b=60$, $B/B_0 = 3$). 
This has significant observational implications since the concentration of the magnetic field and density inside the plasmoids leads to larger numbers of high-$\chi$ particles, and hence greatly enhances photon emissivity there.
In particular, as illustrated in Fig.~3a for our QED case, the local average $<\chi_e>$ can reach significant values ($\sim 0.1$ or higher) in plasmoid cores. Correspondingly, high-energy photon emissivity and energy density are also strongly enhanced at these locations (Fig.~3b). 
Plasmoids thus effectively become brightly shining fireballs [c.f. \citet{Giannios-2013}].

\begin{figure*}[ht!]
\plotone{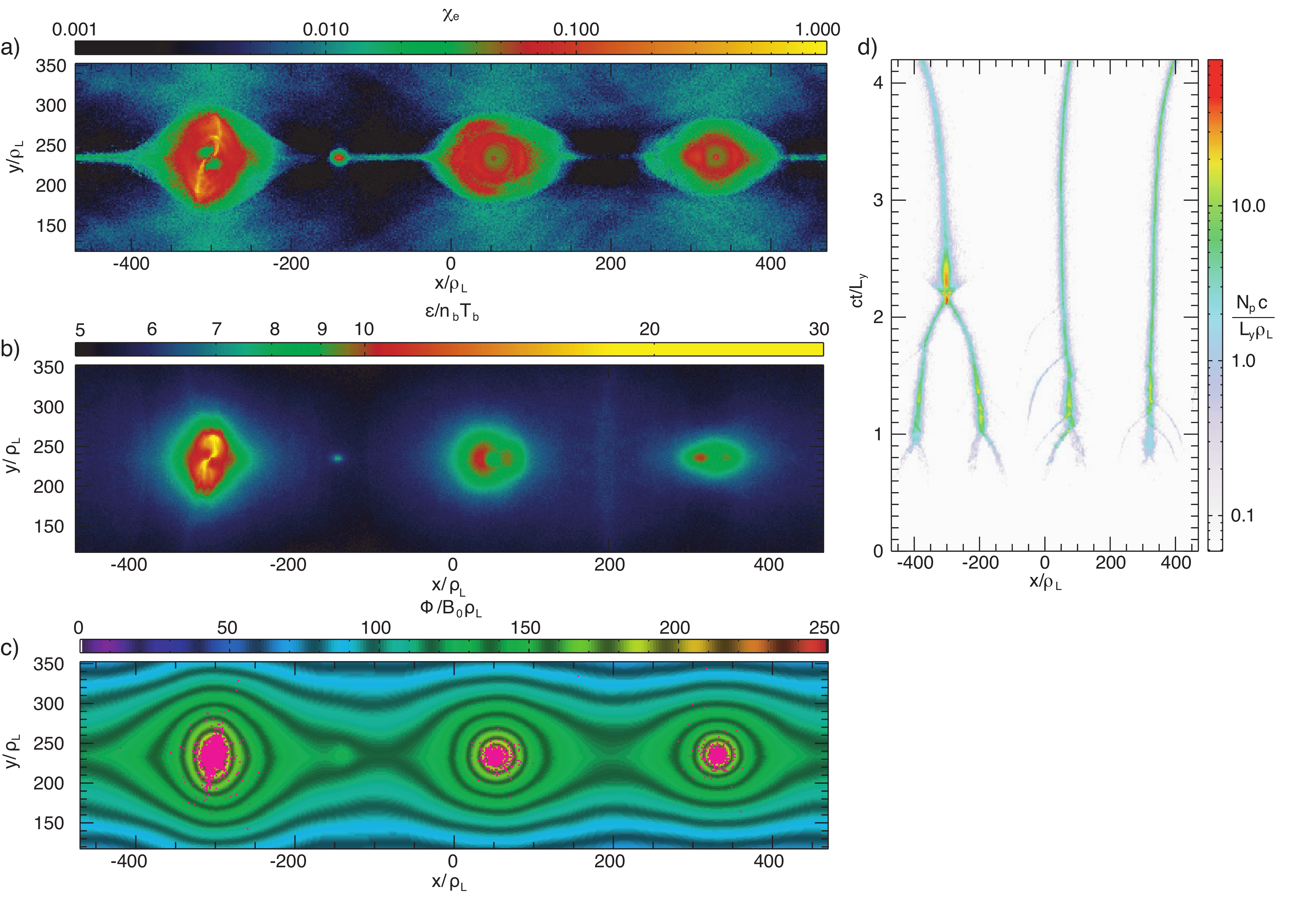}
\caption{Snapshot maps of the local average electron $\chi$ parameter (panel a), of the radiation energy density $\epsilon$ (panel b),
and of the in-plane magnetic flux with the locations of pair production events shown in magenta (panel c) for the QED case at $t = 2.2 L_y/c$ around the current sheet on the upper half of the simulation domain.
Panel d is the space-time map of the pair creation rate density per $c/L_y \rho_L$ averaged over the upper half of the simulation.}
\end{figure*}

\begin{figure*}[ht!]
~~~~~~~~\includegraphics[width=4.1in]{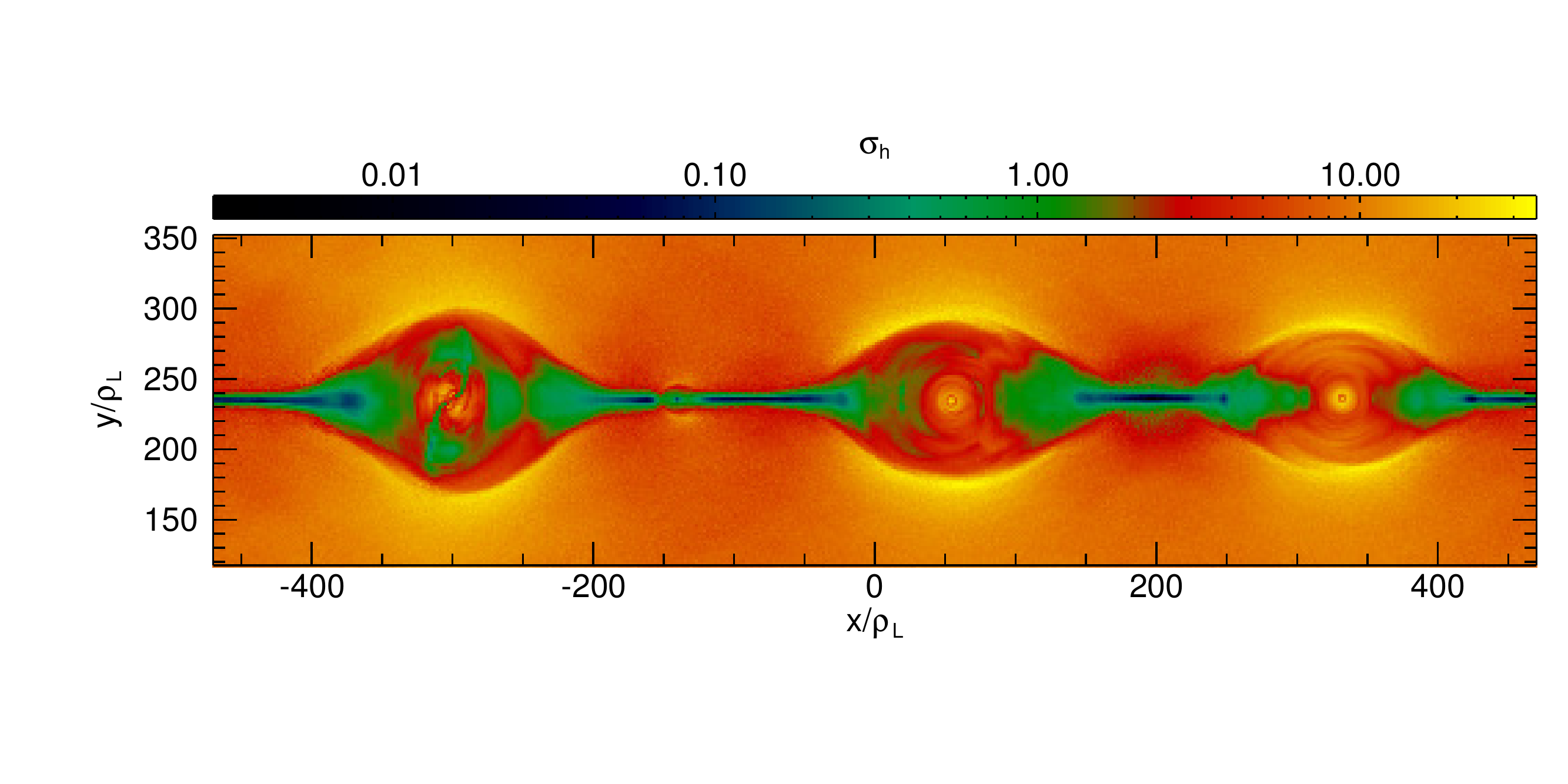}
\caption{Snapshot map of the average $\sigma_h$ parameter, where the local relativistic enthalpy $h_b$ is calculated from the trace of the pressure tensor, for the QED case at $t = 2.2 L_y/c$ around the current sheet in the upper half of the simulation domain.}
\end{figure*}

The spatial coincidence of the local enhancements of gamma-ray photon density and of the magnetic field strength leads to a strong concentration of one-photon QED pair production inside the magnetic islands (Fig.~3c). 
Indeed, using the probabilities given in the Appendix section C.1, one can estimate the characteristic photon decay length $l_{decay}$ --- the distance that a typical hard photon (with $\epsilon_{ph}/m_ec^2 \sim B_Q/B$) travels before producing a pair --- to be $l_{decay} \approx 1700\, c/\Omega_c(B)$ (Appendix section C.2), which corresponds to $(70-15) \rho_L$ for $B = (3-15) B_0$.  
The typical island width in our QED-case simulation (with $B > 3 B_0$) at $t = 1.2 L_y/c $ is $\sim 20 \rho_L$ and grows to $\sim 60 \rho_L$ by $t = 2.2 L_y/c$.  The fact that the island size is larger than $l_{decay}$ allows for the pair production to take place within the island.
Although secondary islands, generated independently of the initial conditions as the inter-plasmoid current layers elongate and themselves become tearing-unstable, are smaller and thus have less photon emission and pair production (most photons leave the small islands before producing pairs), this should not be the case in more realistic, bigger systems where even these secondary islands may grow large enough to exceed the characteristic decay length.  

The $x-t$ diagram [similar to \citet{Nalewajko_etal-2015}] in Fig.~3d shows the location of pair production vs. time and illustrates the creation, motion, and merging of the islands.
We see that both pair production and gamma-ray emission are enhanced at plasmoid mergers.

For the presented QED simulation, the compression of magnetic fields and the strong radiative cooling in the centers of the islands leads to a moderate local $\sigma_h$ comparable to the background~(Fig.~4).
We note that in our simulations with lower density and hence higher $\sigma_h$  (keeping $L_y/\rho_L$ and $B_0/B_Q$ constant; not presented here) the number of produced pairs was increased; significant pair production may thus be expected for such higher $\sigma_h$ systems.

\begin{figure*}[ht!]
\plotone{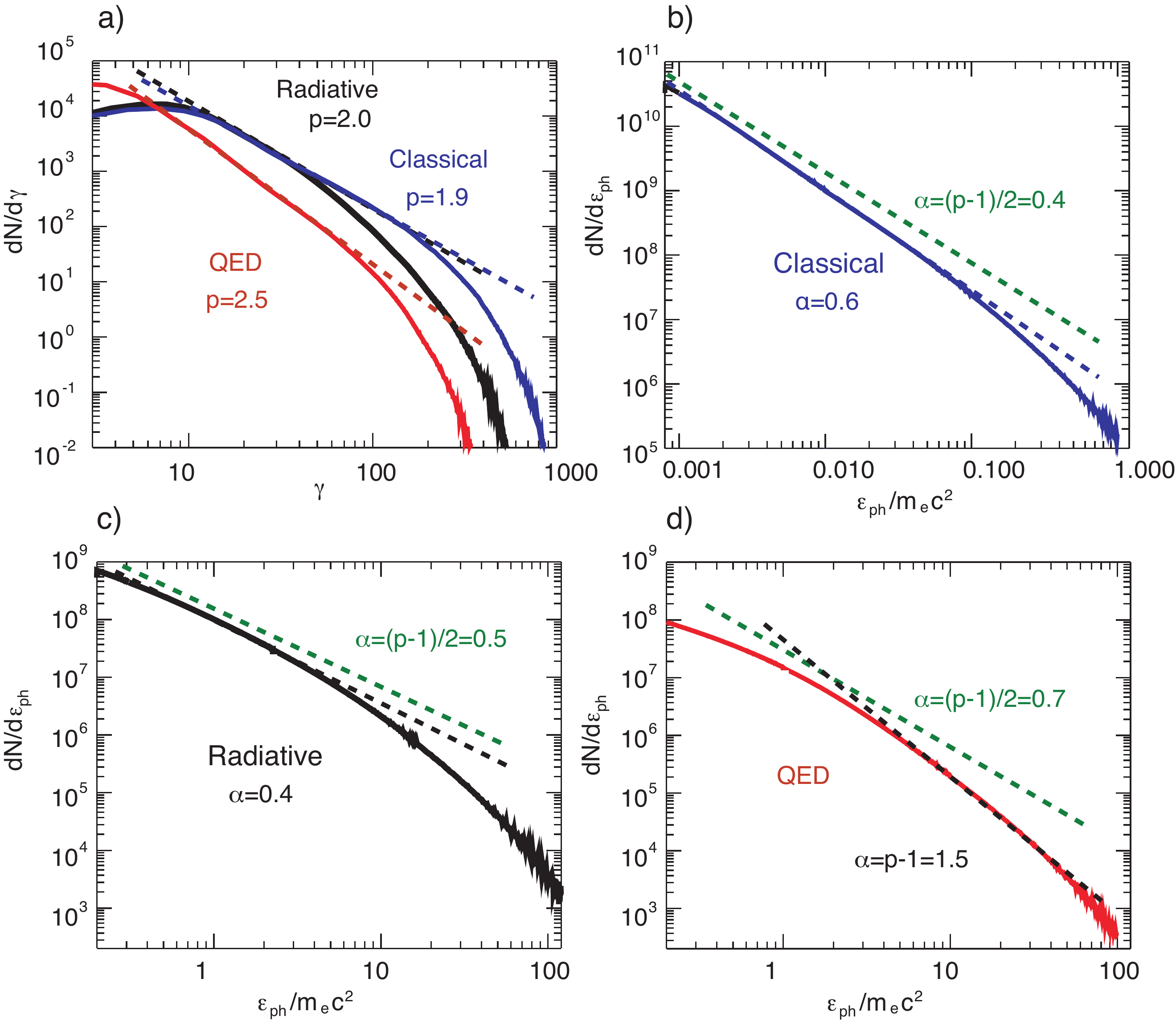}
\caption{Panel (a): Electron energy spectra at $t = 3L_y/c$ for the classical case (blue), the radiative case (black), and the QED case (red). 
The best-fit power-law slopes are shown with the dashed lines. 
The spectra of the photons accumulated in the system by $t = 3L_y/c$ are shown for the classical (panel b), radiative (panel c), and QED (panel d) cases, with best power-law fits represented by the black dashed lines. 
The green dashed lines show the classical synchrotron prediction $\alpha = (p-1)/2$. In the QED case, the best-fit black dashed line is in excellent agreement with the theoretical quantum radiation regime prediction $\alpha = p-1=1.5$.}
\end{figure*}

The main, and often the only, observable signature of reconnection in astrophysical sources is the radiation spectrum, from which the underlying electron energy distribution can be inferred.  All our simulations clearly show nonthermal electron (and positron) acceleration, marked by extended power-law segments, $dN/d\gamma \sim \gamma^{-p}$ (see Fig.~5a). In the classical case, our measured electron spectral index $p \approx 1.8-1.9$ is in agreement with the results of previous non-radiative PIC studies for the given value of $\sigma_h = 6.44$, predicting $p \approx 1.7-2.0$ \citep{Sironi, Guo_etal-2014, Werner_etal-2016, Werner_Uzdensky-2017}.  In the radiative case, energy-dependent radiation cooling steepens the electron spectrum appreciably at the highest energies, while the medium-energy part of the spectrum steepens only slightly  (to $p\approx 2.0$).  Finally, the even stronger radiative cooling in the QED case leads to a significant steepening of the entire spectrum (to $p\approx 2.5$). 
[Note that the presented spectra are all taken at the same fixed time $t = 3 L_y/c$ and integrated over all directions; we leave the investigations of the spectral evolution and of particle and radiation anisotropy to future studies.]

The photon energy spectra, $dN/d\epsilon_{\rm ph}$, are shown in Fig.~4(b-d) for the classical, radiative, and QED cases, respectively, at $t=3L_y/c$. 
For classical synchrotron radiation, a power-law electron spectrum produces a power-law radiation spectrum $\epsilon_{\rm ph}dN/d\epsilon_{\rm ph}\sim \epsilon_{\rm ph}^{-\alpha}$
with $\alpha = (p-1)/2$ [see, e.g. \citet{Rybicki_Lightman-book}].
And indeed, our measured electron and photon spectral indices agree with this relationship reasonably well in both the classical case [$(p-1)/2 \approx 0.4$ vs.~$\alpha\approx 0.6$; Fig.~5b] and in the radiative case [$(p-1)/2 \approx 0.5$ vs.~$\alpha\approx 0.4$; Fig.~5c].  (The modest discrepancy in the classical case is likely because the measured photon spectrum is based on all of the accumulated photons rather than the instantaneous emission spectra.)

In the QED case, however, the measured photon slope ($\alpha \approx 1.5$) is consistently steeper than that predicted classically ($\alpha \approx 0.7$, see Fig.~5d).
The reason for this is that the energy of emitted photons cannot exceed the emitting electron's energy $\gamma m_e c^2$; the critical photon energy thus transitions from the classical synchrotron value $\epsilon_{\rm ph} \sim \gamma^2 m_e c^2 B/B_Q $ to the quantum limit $\epsilon_{\rm ph} = \gamma m_e c^2$. This modifies the $\alpha$ vs.~$p$ relationship from $\alpha = (p-1)/2$ to $\alpha=p-1$, as is confirmed in Fig.~5d (with measured $p\approx 2.5$ and $\alpha \approx 1.5$).
Furthermore, the energy range of the power-law part of the photon spectra can be estimated from the above relations between $\epsilon_{\rm ph}$ and~$\gamma$ and matches with the simulation results.  In the classical case, $B \approx B_0$, and so the electron power-law range of $\gamma = 10 - 200$ yields $\epsilon_{\rm ph}/m_e c^2 \approx 5 \times 10^{-4} - 0.2$ (Fig.~5b).  
In the radiative case, the compression leads to a typical $B \approx 5 B_0$, and so the nonthermal electron energy range of $\gamma = 10 - 40$ translates to $\epsilon_{\rm ph}/m_e c^2 \approx 0.2 - 4$ (Fig.~5c). Finally, in the QED case, where $\epsilon_{\rm ph} \sim \gamma m_e c^2$, the electron power-law range of $\gamma = 6 - 80$ translates directly to $\epsilon_{\rm ph}/m_e c^2 \approx 6 - 80$, clearly visible in Fig.~5d.

\section{Summary} \label{sec:summary}

In summary, we have unambiguously demonstrated, via first-principles PIC simulations that self-consistently incorporate radiation and QED effects, that relativistic reconnection of strong magnetic fields can power intense high-energy radiation flares and lead to pair production. 
We showed that radiation (dominated by synchrotron) cooling and one-photon pair production in strong-field reconnection can lead to remarkable differences from classical relativistic reconnection. 
These effects are greatly enhanced by the cooling-caused compression of the plasma density, pressure, and reconnected magnetic flux inside magnetic islands (plasmoids); the cooling is, in turn, further intensified by the compressed magnetic field.  The resulting powerful emission of gamma-ray photons, in combination with the amplified magnetic field, then leads to enhanced pair production in plasmoid cores. Thus, both high-energy emission and pair creation are strongly concentrated in the plasmoids, effectively turning them into bright, dense and relativistically hot flying fireballs. 
The observable spectra of the emitted radiation are significantly steeper than those produced in classical relativistic reconnection, both because radiation reaction inhibits nonthermal particle acceleration and because of QED effects on the emission from particles with $\chi\sim 1$, resulting in potentially measurable signatures.

%
These results have profound implications for our understanding of the role of reconnection in high-energy astrophysical environments with very strong magnetic fields---most notably, magnetospheres of neutron stars (NS), especially magnetars~\citep{Masada2010}. Our study provides firm support to the hypothesis \citep{Thompson2001,Lyutikov2006, Uzdensky-2011} that magnetic reconnection in the QED regime is capable of powering the spectacular gamma-ray flares observed in a class of magnetars called Soft Gamma Repeaters (SGRs), in which $10^{44}-10^{46}$~ergs is emitted in gamma-rays in just a fraction of a second \citep{Mazets1999,Palmer2005,Turolla2015}.
While our simulations are initialized with thin, intense current sheets (which are necessary for reconnection onset), recent theoretical research has indicated that such structures can indeed form in active magnetar magnetospheres via nonlinear MHD processes similar to those driving the flaring activity in the solar corona. Namely, it is believed that even smooth sheared motions of the magnetic footpoints on a magnetar's surface can drive the force-free field in the magnetosphere above the surface towards explosive development of thinner and thinner current sheets, thus setting the stage for reconnection onset \citep[e.g.,][]{Thompson2002,Uzdensky2002,Parfrey2013}.

Intriguingly, while our study confirms that near-Schwinger-field reconnection readily produces intense gamma-radiation and large numbers of $e^+e^-$ pairs, it also indicates that, due to the concentrated magnetic field enhancement in the plasmoids, strong radiation and significant pair production may take place even in environments with modest ambient magnetic fields, well below the Schwinger field~$B_Q$. This may happen, for example, in the magnetospheres of normal ($10^{11}-10^{12}$~G) NSs (radio- and X-ray pulsars) and in magnetar flares taking place at large distances ($\sim 10$ NS radii) from the star.
Finally, our demonstration of high, order-unity radiative efficiency of reconnection in this parameter regime suggests that prompt (i.e., on the reconnection timescale) radiative cooling is important and needs to be accounted for in NS magnetospheric reconnection \cite[c.f.][]{Uzdensky_Spitkovsky-2014,Philippov_Spitkovsky-2018}.  

Our present investigation opens up exciting new frontiers and lays the groundwork for future studies. First, we envision several important straightforward extensions of the present work: performing more realistic 3D simulations; delving even deeper into the QED reconnection regime (with stronger magnetic fields, i.e., higher~$\sigma_h$); and studying the effects of a guide magnetic field. 

In particular, the extra degree of freedom in 3D would allow the compressed plasma to escape from the pinch equilibrium in plasmoids, making the 2D compression less pronounced. Determining a more realistic upper limit on the compression is thus an open issue requiring 3D simulations.  

Next, in more magnetically-dominated systems (with higher $\sigma_h$, even when $B_0/B_Q$ is kept fixed), yet to be explored, the greater available amount of magnetic energy per particle would lead to stronger heating and nonthermal particle acceleration. The resulting stronger radiative cooling of these energetic particles may then drive further compression of the flux ropes.
Both the enhanced heating/acceleration and the compression of the magnetic field mean a  higher~$\left<\chi_e \right>$, increasing the emission of MeV gamma-rays. Even for modest system sizes, the compactness for one-photon QED pair production may be so high that large numbers of pairs would be created and would eventually trap the radiation and produce an optically thick, hot and dense lepto-photonic fireball, with temperature and density independent of their initial background values \citep{Uzdensky-2011}.

Finally, all these aspects of reconnection are likely to be affected by an out-of-plane guide ($B_z$) magnetic field. A moderate or strong guide field will resist compression and may suppress the efficiency of nonthermal particle acceleration \citep{Werner_Uzdensky-2017}.
At the same time, it can also suppress the relativistic drift-kink instability (RDKI), which develops in the third dimension and competes with the tearing instability \citep{Zenitani_Hoshino-2007}.
The effects of guide  magnetic field are thus another important direction for further study.

Beyond these immediate generalizations, this project paves the way to future rigorous, first-principles exploration of qualitatively new, physically rich regimes of magnetic reconnection and, in fact, of many other relativistic kinetic plasma processes. In these regimes traditional kinetic plasma physics is closely intertwined with radiation, pair creation and annihilation, and perhaps other, even more exotic QED effects.  It thus opens a new research direction --- computational QED plasma astrophysics, which will help unlock the secrets of magnetar flares and other fascinating and exotic astrophysical phenomena.

\acknowledgments

We would also like to thank the anonymous referee for useful comments that improved the paper. This work was supported by the European Research Council (ERC-2015-AdG Grant No. 695008), FCT (Portugal) grant SFRH/IF/01780/2013, DOE grants DE-SC0008409 and DE-SC0008655, and NASA grant NNX16AB28G. DAU gratefully acknowledges the hospitality of the Institute for Advanced Study and the support from the Ambrose Monell Foundation.  Simulations were carried at MareNostrum (Spain) under a PRACE award.

\software{OSIRIS \citep{OSIRIS}}
%

\vspace{5mm}





\appendix
\section{Numerical Setup}

In our simulations, we model a $2L_x \times 2L_y$ domain with two
oppositely directed thin current sheets located at $y = \pm L_y/2$.  The current is
directed out of the $(x,y)$ simulation plane in the respective $\pm z$ directions, which leads to
an asymptotic magnetic field ${\bf B} = B_0 \hat{x}$, between $-L_y > y > L_y$, and
${\bf B} = -B_0 \hat{x}$ on the outside of the two current sheets.  An initially uniform
background Maxwell-J\"uttner population of relativistic electrons and positrons, each with density $n=n_b$ at temperature $T=T_b$, is included to represent the ambient (upstream) plasma. 
This population is initially stationary and does not contribute to the current.  
Furthermore, we include a weak uniform guide magnetic field $B_G = 0.05 B_0$ along the $z$ direction. 

The current and self-consistent magnetic field profiles are in pressure balance in a kinetic equilibrium, known as the relativistic Harris sheet~\citep{Harris1962,KirkHarris}.
The current is carried by counter-drifting Maxwell-J\"uttner distributions of positrons and electrons with a uniform temperature $T_0$, boosted into opposite directions with a uniform velocity $v_d$. 
The lab-frame density profile (of both electrons and positrons) in  the Harris current sheet at $y = \pm L_y/2$ is:
\begin{equation}
	n=\left(n_0 - n_b\right) {\rm sech}^2 \left(\frac{y \mp L_y/2}{\delta}\right),
\end{equation}
here $n_0$ is the total electron (or positron) density at the center of each current sheet.
The self-consistent magnetic field is:
\begin{eqnarray}
	B_x &=&B_0\left[
	-\tanh\left(\frac{y - L_y/2}{\delta}\right)
	+\tanh\left(\frac{y + L_y/2}{\delta}\right)\right.\nonumber\\
	&+&\left.\tanh\left(\frac{y - 3L_y/2}{\delta}\right)
-\tanh\left(\frac{y + 3L_y/2}{\delta}\right) + 1\right].
\end{eqnarray}
We conduct our simulations with periodic boundary conditions, so we also
include the self-consistent magnetic field due to two more current sheets at
$y = 3L_y/2$ and $y = -3L_y/2$ (outside of the simulation box).  This is a
small correction due to the periodic boundary conditions introduced to account for
the exponential tail that passes through the boundary.  In order to facilitate the onset of 
magnetic reconnection, the initial thickness of the current sheet $\delta$ is chosen to
be sufficiently small (of order the gyro-radius of the particles in the sheet), so that the tearing instability growth rate approaches the characteristic cyclotron period~\citep{DaughtonPOP}. 
We normalize all the length scales in our simulation to $\rho_L \equiv \gamma_T m_e c^2/eB_0 = \gamma_T c/\Omega_c$, defined as the Larmor radius of a background particle with a
Lorentz factor corresponding to the peak of the initial upstream relativistic
Maxwell-J\"uttner distribution, $\gamma_T \equiv 2T_b/m_e c^2$, and
choose $\delta > \rho_L$, $\rho_{L0}$, where $\rho_{L0} = \rho_L T_0/T_b$ is
the gyroradius of a typical particle in the current sheet.

The three main physical parameters that describe the upstream plasma conditions outside of the current sheets---$T_b$, $n_b$, and $B_0$---define two important dimensionless parameters: the  magnetization $\sigma_h$ and the plasma-$\beta$ parameter, $\beta_{up}$ (the ratio of the background plasma pressure to the magnetic pressure):
\begin{eqnarray}
	\label{eq-sigma_h}
    \sigma_h &\equiv &\frac{B^2}{4 \pi (2n_b)h_b}, \\
	\label{eq-beta_up}
    \beta_{up} & \equiv & \frac{8 \pi (2n_b) T_b}{B_0^2} = \frac{2 T_b}{h_b} \frac{1}{\sigma_h}.  
\end{eqnarray}

The subscript $h$ refers to the "hot"magnetization $\sigma_h$, defined with the upstream background relativistic enthalpy per particle~$h_b$~\citep{Melzani}. 
In the nonrelativistic limit ($T_b \ll m_e c^2$), the enthalpy $h_b \approx m_e c^2 +  5/2 T_{b}$ is dominated by the rest-mass $m_e c^2$  and so the "hot" magnetization $\sigma_h$ approaches the so-called "cold" magnetization $\sigma_c \equiv B_0^2/4 \pi (2n_b) m_e c^2$, which is often used in the literature.  In the ultrarelativistic limit ($T_b \gg m_e c^2$), however, $h_b \approx 4T_{b}$, and then $\sigma_h = 1/(2\beta_{up})$.

Using the $\beta_{up}$ parameter allows us to cast the electron and positron drift speed inside the two Harris current layers, determined by Amp\`ere's law, in a convenient form as
\begin{equation}
	\label{amperes}
	\frac{v_d}{c} = \frac{1}{\beta_{up}}\frac{\rho_L}{\delta}\frac{n_b}{n_0-n_b}. 
\end{equation}
In addition, the temperature $T_0$ of the drifting plasma in the layer, determined by the cross-layer pressure balance, can be written as 
\begin{equation}
	\label{pressurebalance1}
	\frac{T_0}{m_e c^2} = \frac{T_b}{m_e c^2}\frac{\gamma_d}{\beta_{up}}\frac{n_b}{n_0- n_b}, 
\end{equation}
where $\gamma_d \equiv 1/\sqrt{1- v_d^2/c^2}$.

\section{Conditions of Applicability of the Model} \label{sec-conditions}
\setcounter{equation}{0}
The physical parameters needed to be specified for magnetic reconnection starting from a Harris sheet~\citep{Harris1962} equilibrium are the following:
(A) ambient (upstream) pair-plasma parameters: 
the background electron/positron density $n_b$, the background temperature~$T_b$, the upstream (reconnecting) magnetic field~$B_0$, and the out-of-plane guide field~$B_G$; 
(B) Initial current-layer parameters: the electron density in the center of the current sheet~$n_0$, and the current half-thickness~$\delta$. (The temperature and the drift velocity can then be determined by force balance and Amp\`ere's law, see the Numerical Setup.)
(C) The system's dimensions $L_x$ and $L_y$, which set the typical time of reconnection; in the collisionless case considered here, it is several light crossing times $t_{cross} = L_y/c$. 

We believe that, as long as the system size is large enough so that the overall number of background particles dominates over the drifting population, $n_b L_y \gg n_0 \delta$, the exact values of the initial current-layer parameters (parameter group B above) are not critically important and affect only the initial transient stage of reconnection. 
In contrast, the initial background plasma parameters ($T_b$, $n_b$, and $B_0$, i.e., parameter group A) are fundamentally important as they determine the two key dimensionless parameters, $\sigma_h$ and $\beta_{up}$ [see Eqs.~(\ref{eq-sigma_h})-(\ref{eq-beta_up})], which control the reconnection regime. 
[The system size $L_x=L_y$ (group C) is also important as it needs to be large enough for the reconnection process to proceed in the large-system, plasmoid-dominated regime.]
It is thus important to describe our reasons for choosing the specific values of these parameters for our study.  Our choices are dictated in part by the considerations of simplicity and computational feasibility (which, for example, limit the maximum system size that we can achieve) and in part by various physical assumptions and validity conditions for our model, which we discuss in this section. 

For clarity, we present these conditions on the 2D $(n_b,T_b)$ parameter-space map shown in Fig.~A.~1. We show several lines delineating the regions where certain additional physical processes that we do not include become non-negligible. These lines represent the most restrictive constraints and are applied to both the background and the Harris populations, characterized by their values of $T$ (relativistic or non-relativistic), $n$, and~$B_0$.

The physical requirements are the following:
\begin{itemize}
	\item the density does not surpass the Compton density $n_C \equiv (\hbar/m_ec)^{-3} = 1.739 \times 10^{31}\, {\textrm{cm}^{-3}}$;
    \item the relativistic field invariants are small ($|E^2-B^2|/B_Q^2,|E\cdot B|/B_Q^2 \ll 1$);
    this is part of the constant cross-field approximation, used in determining our photon emission/ pair production rates
    (in red);
	\item no strong upstream cooling ($t_{rad}, t_{rad,r} \gg t_{cross}$); the background plasma does not cool significantly during the crossing time of the system;  
        here we define $t_{rad}$ and $t_{rad,r}$ as the characteristic cyclo-synchrotron cooling times for the nonrelativistic and ultrarelativistic cases, respectively:
\begin{equation}
t_{rad} =\frac{3}{4}\frac{1}{\alpha_{\rm fs}}\frac{B_Q}{B}\frac{1}{\Omega_c},~~~
t_{rad,r} =\frac{3}{2}\frac{1}{\alpha_{\rm fs}\gamma_T}\frac{B_Q}{B}\frac{1}{\Omega_c}
\end{equation}
where $\alpha_{\rm fs}\equiv e^2/\hbar c$ is the fine structure constant
(in blue);
	\item quantum degeneracy effects can be neglected [the temperature is high compared to the Fermi energy $E_F \equiv \hbar c (3 \pi^2 n)^{1/3}$] (in purple);
	\item cyclotron orbits are not quantized [the temperature is high compared to Landau energy levels $\sim \hbar \Omega_c$] (in cyan);
	\item collective effects dominate [large plasma parameter
		$\Lambda \equiv n \lambda_D^{3}$, where $\lambda_D$ is the Debye length] (in green);
	\item collisionless plasma (a typical particle does not collide during
		the light crossing time of the system $t_{cross}= L_y/c \ll \nu^{-1}$).
        The electron-electron and electron-positron collision rate is $\nu \sim
		\omega_{pe}\textrm{ln}\left(\Lambda_C\right)/ \Lambda$, where
		$\omega_{pe}$ is the classical plasma frequency, and $\textrm{ln}\left(\Lambda_C\right)$ is the Coulomb logarithm (in dark green).
\end{itemize}

The parameter space is shown in Fig.~A.~1, where all these conditions are met in the white region, bounded above by the highly radiative regime in blue, and below by the collisional regime in green. Specifically, in order to be able to cast these conditions in the ($n_b,T_b$) parameter space, we adopted a fixed value $\beta_{up}=0.0776$ for all our simulations; this value is chosen to be small compared to unity so that the upstream region is magnetically dominated. In addition, we set $n_0/n_b = 10$, $\delta/\rho_L = 2.55$, and $L_y/\rho_L=472$ for all the runs.  These parameters yield $T_0/m_ec^2=6.92$, $v_d/c=0.56$ ($\gamma_d=1.21$), and $\delta/\rho_{L0} = 1.47$ where $\rho_{L0}$ is the Larmor radius based on~$T_0$. We have thus chosen $\delta$ larger than, but close enough to $\rho_{L0}$ so that tearing commences quickly. 
The specific three simulations presented in this paper are indicated in Fig.~A.~1 by the red circles; they all correspond to the same initial background temperature $T_b/m_ec^2=4.0$, while the background density is varied, $n_b = 1.90 \times 10^{19}, 1.90 \times 10^{23}, 1.90 \times 10^{25}$ cm$^{-3}$ (equivalent to varying $B/B_Q = 4.53 \times 10^{-6},4.53 \times 10^{-4}, 4.53 \times 10^{-3}$).

We performed our simulations taking advantage of the OSIRIS framework~\citep{OSIRIS} with $3840 \times 3840$ computational cells of size $\Delta x = \Delta y =\rho_L/4$, initially with $16$ particles per species in each cell. The presented simulations are run for 4.2 light crossing times~$L_y/c$, with a time step of $\Delta t = 0.142 \rho_L/c = 0.142 \gamma_T \Omega_c^{-1}$.

\renewcommand{\thefigure}{A\arabic{figure}}
\setcounter{figure}{0}
\begin{figure*}[ht!]
\plotone{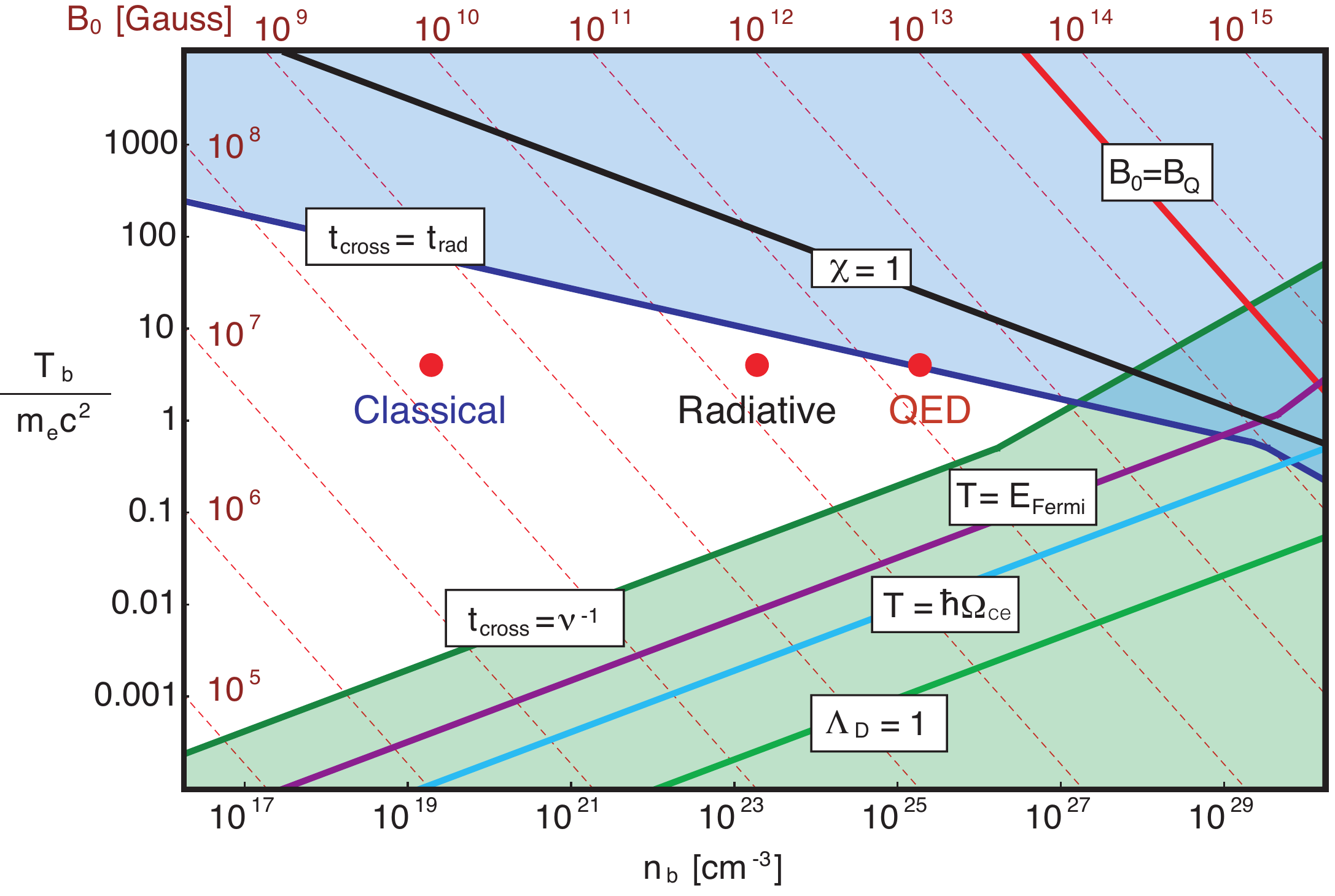}
\caption{The parameter space of $T_b$ and $n_b$ keeping $\beta_{up} = 0.0776$,
$\delta/\rho_L = 2.55$, $n_0/n_b = 10$, and $L_y/\rho_L = 472$ constant.  The blue region above represents the highly radiative regime, where $t_{cross} > t_{rad}$, and the green region represents the highly collisional regime where $t_{cross} > \nu^{-1}$. 
The boundaries of the other physical requirements on our
assumptions lie in either the radiative or the collisional regimes.
The black line corresponds to $\left<\chi_e\right> = 1$ for the thermal particles of the initial setup. 
In each of the lines the temperature and density are taken from the population --- either Harris ($T_0$, $n_0$) or background ($T_b$, $n_b$)--- that leads to the most restrictive limits, using the appropriate relativistic or non-relativistic expressions based on the values of $T_0$ and~$T_b$.
Levels of constant magnetic field are indicated by thin dashed red lines and the red circles show the three simulations reported in this paper.}
\end{figure*}

We also show that the relativistic field invariants remain small as the system evolves in Fig~.A.~2.
\begin{figure*}[ht!]
+\plotone{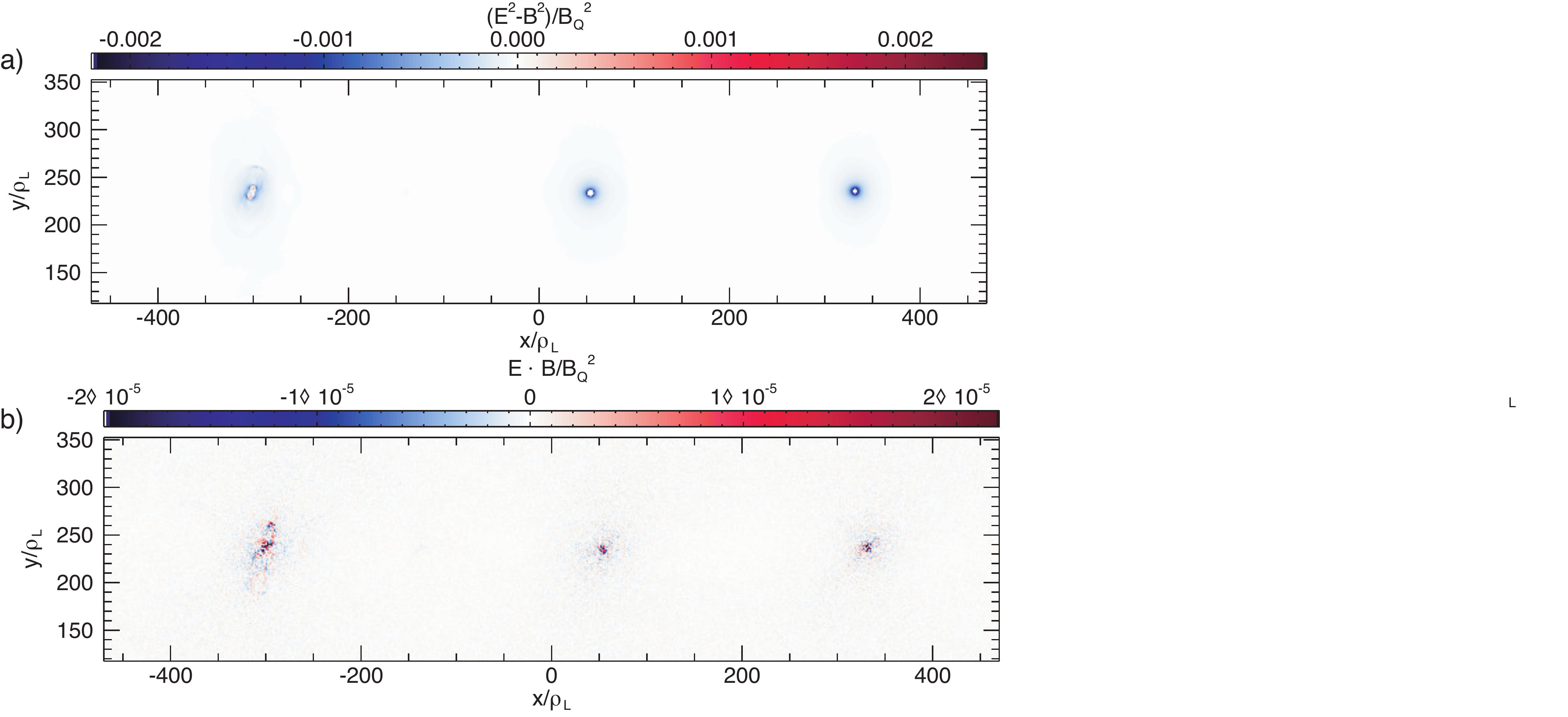}
\caption{The maps of the relativistic invariants $(E^2-B^2)/B_Q^2$ (panel a) and $E\cdot B/B_Q^2$ (panel b)
at $t = 2.2 L_y/c$ for the QED case around the current sheet on the upper half of the simulation domain.}
\end{figure*}

\section{QED Processes} \label{sec:rotate}
\setcounter{equation}{0}
\subsection{Probability rates}
Rigorous investigation of some so-far unexplored reconnection regimes must take into account various QED processes associated with strong magnetic fields.  
Many such processes can in principle take place, but in this work, we only consider two quantum processes that have the highest probabilities. 
These processes (implemented in our code through a Monte-Carlo module in the particle-in-cell loop) are 
(1) single photon emission due to non-linear Compton scattering in intense electromagnetic fields 
(with self-consistent back-reaction recoil on the emitting electron/positron), which is a QED extension of the classical synchrotron radiation; 
and 
(2) single-photon pair creation from the decay of a hard gamma-ray photon ($\hbar\omega > 2 m_e c^2$) in intense electromagnetic fields, also known as the Breit-Wheeler pair production process~\citep{Ritus_thesis}.
Other possible quantum processes such as photon splitting, Compton scattering, two-photon pair creation, and pair annihilation can in principle occur. Photon splitting is only relevant for $B \gtrsim B_Q$, whereas the other processes have cross-sections that are at best $\sim r_e^2$, where $r_e$ is the classical electron radius. The ratio between the mean free path of a particle before experiencing one of the simulated processes $\lambda$, and the other processes $\lambda_\sigma$, is:
	\begin{equation}
		\frac{\lambda}{\lambda_\sigma} \sim \alpha_{fs} \frac{n}{n_C}\frac{B_Q}{B},
	\end{equation}
where $n$ is the density of the species the particle will interact with; this ratio is much smaller than~$1$ in all the regimes that we consider.
 The respective probability rates for photon emission and pair creation depend on the invariant quantum parameter $\chi$ and the energy of the particle. 

	The $\chi$ parameter determines if classical or QED interactions dominate the physics and is defined
	using the 4-momentum $p^\mu$ of the particle (electron/positron, or photon):
	\begin{equation}\label{chif}
		\chi=\frac{\sqrt{(p_\mu F^{\mu\nu})^2}}{B_Q~m_ec}.
	\end{equation}

The parameter $\chi$ will be denoted as $\chi_e$ for electrons, and $\chi_\gamma$ for photons where $p_\mu = \hbar k_\mu$, and $k_\mu$ is the photon wave 4-vector. 
We can also express $\chi$ as a function of 3-vectors and the background electric and magnetic field vectors: 
	\begin{equation}\label{chi_3vectordef}
		\chi=\frac{1}{B_Q}\sqrt{\left(\gamma \vec{E}+\frac{\vec{p}}{mc}\times \vec{B} \right)^2 - \left( \frac{\vec{p}}{mc}\cdot \vec{E} \right)^2}.
	\end{equation}
    $\gamma = p_0 = \epsilon_e/m_ec^2 \rightarrow p_0=\epsilon_\gamma/m_ec^2$ for photons,	where $\epsilon_e$ is the electron energy, and $\epsilon_\gamma=\hbar \omega$ is the photon energy.
The differential probability rate of photon emission with $\chi_\gamma$ by nonlinear Compton scattering of an electron with $\chi_e$ is then given~\citep{Ritus_thesis} by

\begin{equation}\label{compton_rate}
		\frac{d^2 P}{dt~d\chi_\gamma}=\frac{\alpha_{\rm fs}}{\sqrt{3}\pi t_C \gamma \chi_e} 
		\left[ \left( 1 - \xi +\frac{1}{1-\xi} \right) K_{2/3}(\tilde{\chi}) - \int^\infty _{\tilde{\chi}}  dx K_{1/3}(x)  
		\right],
	\end{equation}
    where $t_C \equiv \hbar/m_ec^2$ is the Compton time,
	$\tilde{\chi}=2\xi/(3\chi_e(1-\xi))$, $\xi=\chi_\gamma/\chi_e$, and $K_\alpha(x)$ is the modified Bessel function of the second kind. Integrating Eq.~(\ref{compton_rate}) over $\chi_\gamma$ results in the likely number of photons that would be emitted per unit time (essentially in the direction of the emitting particle's momentum in accordance with the limiting case of relativistic beaming where $\gamma \rightarrow \infty$), 
        \begin{eqnarray}
		 \label{compton_rate_limits}
		 \frac{dP}{dt}&=&\int d\chi_\gamma \frac{d^2 P}{dt~ d\chi_\gamma} \nonumber \\ 
					    &\approx&  1.46\, \frac{\alpha_{\rm fs}}{t_C \gamma}\chi_e ^{2/3}~~~~\textrm{ for }  \chi_e\gg1 \nonumber\\
			      &\approx& 1.44\,\frac{\alpha_{\rm fs}}{t_C \gamma}\chi_e~~~~~~~\textrm{ for }   \chi_e \ll 1 .
	 \end{eqnarray}
The total radiated power is 
	\begin{equation}\label{prad_qed}
		P_{rad}=\int d\epsilon_\gamma~\epsilon_\gamma \frac{d^2 P}{dt~d\epsilon_\gamma}= 
		\frac{\epsilon_e}{\chi_e} \int d\chi_\gamma~\chi_\gamma \frac{d^2 P}{dt~d\chi_\gamma},
	\end{equation}
	assuming $\gamma \gg 1$ and thus $\chi_e/\chi_\gamma = \epsilon_e/\epsilon_\gamma$. For $\xi\ll1$ (valid for nearly all photons if $\chi_e \ll 1$), $P_{rad}$ given by Eq.~(\ref{prad_qed}) reduces to the classical synchrotron radiated power:
 \begin{equation}
        \label{prad_classical}
		P_{rad}=\frac{2}{3}\frac{e^2}{c}\gamma^2\Omega_c^2\sin^2\alpha
        =\frac{2}{3}\frac{\alpha_{\rm fs} m_e c^2}{t_C}\chi_e^2,  	
 \end{equation}
        where $\alpha$ is the pitch angle of the radiating particle.
        In our simulations, the emitted photons are treated as computational particles that are propagated through the simulation domain along straight lines but have some probability of decaying into pairs.
	The differential rate of pair production of an electron and a positron with $\chi_e$ by a photon with $\chi_\gamma$ in a background electromagnetic field is given~\citep{Ritus_thesis} by

\begin{equation}\label{pair_rate}
		\frac{d^2 P}{dt~d\chi_e}=\frac{\alpha_{\rm fs} m_e c^2}{\sqrt{3}\pi t_C \epsilon_\gamma \chi_\gamma} 
		\left[ \left( \frac{\xi^+}{\xi^-} +\frac{\xi^-}{\xi^+} \right) K_{2/3}(\tilde{\chi}) + \int^\infty _{\tilde{\chi}}  dx K_{1/3}(x)  \right]
	\end{equation} 
	where $\tilde{\chi}=2/(3 \chi_\gamma~ \xi^+\xi^-)$ and $\xi^+=\chi_e/\chi_\gamma=1-\xi^-$. The total rate for this process can be approximated for very small or very high $\chi_\gamma$ in the following way: 
	\begin{eqnarray}
		 \label{eq:pairrate}
		 \frac{dP}{dt}&=&\int d\chi_e \frac{d^2 P}{dt~ d\chi_e} \nonumber \\ 
					    &\approx&  0.38\,\frac{\alpha_{\rm fs} m_ec^2}{t_C \epsilon_\gamma}\chi_\gamma ^{2/3}~~~~~~~~~~~~~~~~~~~~~\textrm{ for }  \chi_\gamma\gg1 \nonumber\\
			      &\approx& 0.23\,\frac{\alpha_{\rm fs} m_ec^2}{t_C \epsilon_\gamma}\chi_\gamma\exp \left( -\frac{8}{3\chi_\gamma} \right)\textrm{ for }   \chi_\gamma \ll 1 .
	 \end{eqnarray}

The leptons (i.e., electrons and positrons) are divided into two categories; mildly relativistic particles ($\gamma < 10$) and ultrarelativistic particles ($\gamma \geq 10$). This division is not {\it ad hoc}, it is based on the fact that the above QED probabilities are derived in the limit $\gamma \gg 1$. 

The leptons in the first category ($\gamma < 10$) have $\chi \ll 1$ and thus the radiation-reaction force on them can be described using the classical relativistic Landau-Lifshitz formula~\citep{LandauLifshitz}. The energy lost to radiation is calculated using the Larmor formula and the total radiated energy is recorded as a function of time.

In the second category, when $\gamma \geq 10$, the leptons emit discrete photons according to the aforementioned QED probabilities. When a photon is emitted, the recoil is self-consistently implemented using the conservation of momentum. Unfortunately, due to memory constraints, we cannot keep track of all the photons emitted on the grid. In our simulations, we only track photons above a certain energy~$\epsilon_{cut}$.
We choose this cutoff either as $\epsilon_{cut}= 2m_e c^2$, the minimum energy for a photon that could potentially produce a pair, or as the lower end of the gamma-ray spectra we wish to plot (which was used for the three simulations presented).

\subsection{Photon decay length}
In this section we justify the expression used in the manuscript for the decay length of a hard photon [$l_{decay} = 1700 c/\Omega_c(B)$].
A very good approximation for the pair production rate [see Eq.~(\ref{eq:pairrate})], over the full range of $\chi_\gamma$ is~\citep{Erber}:
\begin{eqnarray}
	\frac{dP}{dt} &=& \frac{4}{25} \frac{\alpha_{\rm fs}}{t_C} \frac{m_ec^2}{\epsilon_\gamma} K_{1/3}^2\left(\frac{4}{3\chi_\gamma}\right)\nonumber \\
	              &=& \frac{4}{25} \frac{\alpha_{\rm fs} \Omega_c}{\chi_\gamma} K_{1/3}^2\left(\frac{4}{3\chi_\gamma}\right)\nonumber \\
			     &\sim& 6\times10^{-4} \Omega_c(B)~~~~\textrm{ for }  3 < \chi_\gamma < 100 ,
\end{eqnarray}
assuming the photons move perpendicular to the magnetic field.
Now
\begin{equation}
	l_{decay} \equiv \frac{c}{dP/dt} \approx 1700\frac{c}{\Omega_c(B)}.
\end{equation}
As long as $\chi_\gamma > 3$, i.e. the hard photon has $\epsilon_{\gamma}/m_ec^2 \sim B_Q/B$, this approximate decay length is valid.
Note that for large $\chi_\gamma > 100$, $l_{decay}$ increases as $\chi_\gamma^{1/3}$.


\begin{thebibliography}{}
\expandafter\ifx\csname natexlab\endcsname\relax\def\natexlab#1{#1}\fi
\providecommand{\url}[1]{\href{#1}{#1}}

\bibitem[{{Beloborodov}(2017)}]{Beloborodov-2017}
{Beloborodov}, A.~M. 2017, \apj, 850, 141

\bibitem[{Bennett(1934)}]{Bennett}
Bennett, W.~H. 1934, Phys. Rev., 45, 890

\bibitem[{{Cerutti} {et~al.}(2016){Cerutti}, {Philippov}, \&
  {Spitkovsky}}]{Cerutti_etal-2016}
{Cerutti}, B., {Philippov}, A.~A., \& {Spitkovsky}, A. 2016, Monthly Notices of
  the Royal Astronomical Society, 457, 2401

\bibitem[{Cerutti {et~al.}(2013)Cerutti, Werner, Uzdensky, \&
  Begelman}]{Cerutti2013}
Cerutti, B., Werner, G.~R., Uzdensky, D.~A., \& Begelman, M.~C. 2013, The
  Astrophysical Journal, 770, 147

\bibitem[{Cerutti {et~al.}(2014)Cerutti, Werner, Uzdensky, \&
  Begelman}]{Cerutti2014}
---. 2014, The Astrophysical Journal, 782, 104

\bibitem[{{Daughton}(1999)}]{DaughtonPOP}
{Daughton}, W. 1999, Physics of Plasmas, 6, 1329

\bibitem[{Erber(1966)}]{Erber}
Erber, T. 1966, Rev. Mod. Phys., 38, 626

\bibitem[{Fonseca {et~al.}(2002)Fonseca, Silva, Tsung, Decyk, Lu, Ren, Mori,
  Deng, Lee, Katsouleas, \& Adam}]{OSIRIS}
Fonseca, R.~A., Silva, L.~O., Tsung, F.~S., {et~al.} 2002, {OSIRIS: A
  three-dimensional, fully relativistic particle in cell code for modeling
  plasma based accelerators}, Vol. 2331 (Springer Berlin / Heidelberg),
  342--351

\bibitem[{{Giannios}(2013)}]{Giannios-2013}
{Giannios}, D. 2013, Monthly Notices of the Royal Astronomical Society, 431,
  355

\bibitem[{Grismayer {et~al.}(2016)Grismayer, Vranic, Martins, Fonseca, \&
  Silva}]{GrismayerPOP}
Grismayer, T., Vranic, M., Martins, J.~L., Fonseca, R.~A., \& Silva, L.~O.
  2016, Physics of Plasmas, 23

\bibitem[{Grismayer {et~al.}(2017)Grismayer, Vranic, Martins, Fonseca, \&
  Silva}]{GrismayerPRE}
---. 2017, Phys. Rev. E, 95, 023210

\bibitem[{{Guo} {et~al.}(2014){Guo}, {Li}, {Daughton}, \&
  {Liu}}]{Guo_etal-2014}
{Guo}, F., {Li}, H., {Daughton}, W., \& {Liu}, Y.-H. 2014, Physical Review
  Letters, 113, 155005

\bibitem[{Harris(1962)}]{Harris1962}
Harris, E.~G. 1962, Il Nuovo Cimento (1955-1965), 23, 115

\bibitem[{Jaroschek \& Hoshino(2009)}]{Hoshino}
Jaroschek, C.~H., \& Hoshino, M. 2009, Phys. Rev. Lett., 103, 075002

\bibitem[{{Kaspi} \& {Beloborodov}(2017)}]{Kaspi2017}
{Kaspi}, V.~M., \& {Beloborodov}, A.~M. 2017, Annual Review of Astronomy and
  Astrophysics, 55, 261

\bibitem[{Kirk \& Skj{\ae}raasen(2003)}]{KirkHarris}
Kirk, J.~G., \& Skj{\ae}raasen, O. 2003, The Astrophysical Journal, 591, 366

\bibitem[{Klepikov(1954)}]{Klepikov}
Klepikov, N.~P. 1954, Zhur. Esptl. i Teoret. Fiz., 26

\bibitem[{{Landau} \& {Lifshitz}(1975)}]{LandauLifshitz}
{Landau}, L.~D., \& {Lifshitz}, E.~M. 1975, {The classical theory of fields}
  (Oxford: Pergamon Press)

\bibitem[{{Lyubarskii}(1996)}]{Lyubarsky-1996}
{Lyubarskii}, Y.~E. 1996, Astronomy \& Astrophysics, 311, 172

\bibitem[{Lyubarsky \& Kirk(2001)}]{Lyubarsky}
Lyubarsky, Y., \& Kirk, J.~G. 2001, The Astrophysical Journal, 547, 437

\bibitem[{{Lyutikov}(2006)}]{Lyutikov2006}
{Lyutikov}, M. 2006, Monthly Notices of the Royal Astronomical Society, 367,
  1594

\bibitem[{{Masada} {et~al.}(2010){Masada}, {Nagataki}, {Shibata}, \&
  {Terasawa}}]{Masada2010}
{Masada}, Y., {Nagataki}, S., {Shibata}, K., \& {Terasawa}, T. 2010,
  Publications of the Astronomical Society of Japan, 62, 1093

\bibitem[{{Mazets} {et~al.}(1999){Mazets}, {Cline}, {Aptekar'}, {Butterworth},
  {Frederiks}, {Golenetskii}, {Il'Inskii}, \& {Pal'Shin}}]{Mazets1999}
{Mazets}, E.~P., {Cline}, T.~L., {Aptekar'}, R.~L., {et~al.} 1999, Astronomy
  Letters, 25, 628

\bibitem[{{McKinney} \& {Uzdensky}(2012)}]{McKinney_Uzdensky-2012}
{McKinney}, J.~C., \& {Uzdensky}, D.~A. 2012, \mnras, 419, 573

\bibitem[{{Melzani} {et~al.}(2013){Melzani}, {Winisdoerffer}, {Walder},
  {Folini}, {Favre}, {Krastanov}, \& {Messmer}}]{Melzani}
{Melzani}, M., {Winisdoerffer}, C., {Walder}, R., {et~al.} 2013, Astron.
  Astrophys., 558, A133

\bibitem[{{Nalewajko} {et~al.}(2015){Nalewajko}, {Uzdensky}, {Cerutti},
  {Werner}, \& {Begelman}}]{Nalewajko_etal-2015}
{Nalewajko}, K., {Uzdensky}, D.~A., {Cerutti}, B., {Werner}, G.~R., \&
  {Begelman}, M.~C. 2015, The Astrophysical Journal, 815, 101

\bibitem[{Palmer {et~al.}(2005)Palmer, Barthelmy, Gehrels, Kippen, Cayton,
  Kouveliotou, Eichler, Wijers, Woods, Granot, Lyubarsky, Ramirez-Ruiz,
  Barbier, Chester, Cummings, Fenimore, Finger, Gaensler, Hullinger, Krimm,
  Markwardt, Nousek, Parsons, Patel, Sakamoto, Sato, M., \& J.}]{Palmer2005}
Palmer, D.~M., Barthelmy, S., Gehrels, N., {et~al.} 2005, Nature, 434,
  1107Ð1109

\bibitem[{{Parfrey} {et~al.}(2013){Parfrey}, {Beloborodov}, \&
  {Hui}}]{Parfrey2013}
{Parfrey}, K., {Beloborodov}, A.~M., \& {Hui}, L. 2013, The Astrophysical
  Journal, 774, 92

\bibitem[{{Philippov} \& {Spitkovsky}(2018)}]{Philippov_Spitkovsky-2018}
{Philippov}, A.~A., \& {Spitkovsky}, A. 2018, \apj, 855, 94

\bibitem[{Ritus(1985)}]{Ritus_thesis}
Ritus, V. 1985, Journal of Soviet Laser Research, 6, 497

\bibitem[{{Rybicki} \& {Lightman}(1979)}]{Rybicki_Lightman-book}
{Rybicki}, G.~B., \& {Lightman}, A.~P. 1979, {Radiative processes in
  astrophysics} (Wiley-Interscience, New York)

\bibitem[{Sironi {et~al.}(2016)Sironi, Giannios, \& Petropoulou}]{Sironi2006}
Sironi, L., Giannios, D., \& Petropoulou, M. 2016, Monthly Notices of the Royal
  Astronomical Society, 462, 48

\bibitem[{Sironi \& Spitkovsky(2014)}]{Sironi}
Sironi, L., \& Spitkovsky, A. 2014, The Astrophysical Journal Letters, 783, L21

\bibitem[{{Thompson}(1994)}]{Thompson}
{Thompson}, C. 1994, Monthly Notices of the Royal Astronomical Society, 270,
  480

\bibitem[{Thompson \& Duncan(2001)}]{Thompson2001}
Thompson, C., \& Duncan, R.~C. 2001, The Astrophysical Journal, 561, 980

\bibitem[{{Thompson} {et~al.}(2002){Thompson}, {Lyutikov}, \&
  {Kulkarni}}]{Thompson2002}
{Thompson}, C., {Lyutikov}, M., \& {Kulkarni}, S.~R. 2002, The Astrophysical
  Journal, 574, 332

\bibitem[{Turolla {et~al.}(2015)Turolla, Zane, \& Watts}]{Turolla2015}
Turolla, R., Zane, S., \& Watts, A.~L. 2015, Reports on Progress in Physics,
  78, 116901

\bibitem[{{Uzdensky}(2002)}]{Uzdensky2002}
{Uzdensky}, D.~A. 2002, The Astrophysical Journal, 572, 432

\bibitem[{Uzdensky(2011)}]{Uzdensky-2011}
Uzdensky, D.~A. 2011, Space Science Reviews, 160, 45

\bibitem[{{Uzdensky}(2016)}]{Uzdensky-2016}
{Uzdensky}, D.~A. 2016, Magnetic Reconnection: Concepts and Applications, 427,
  473

\bibitem[{{Uzdensky} \& {Spitkovsky}(2014)}]{Uzdensky_Spitkovsky-2014}
{Uzdensky}, D.~A., \& {Spitkovsky}, A. 2014, The Astrophysical Journal, 780, 3

\bibitem[{{Werner} \& {Uzdensky}(2017)}]{Werner_Uzdensky-2017}
{Werner}, G.~R., \& {Uzdensky}, D.~A. 2017, The Astrophysical Journal Letters,
  843, L27

\bibitem[{{Werner} {et~al.}(2016){Werner}, {Uzdensky}, {Cerutti}, {Nalewajko},
  \& {Begelman}}]{Werner_etal-2016}
{Werner}, G.~R., {Uzdensky}, D.~A., {Cerutti}, B., {Nalewajko}, K., \&
  {Begelman}, M.~C. 2016, The Astrophysical Journal Letters, 816, L8

\bibitem[{{Zenitani} \& {Hoshino}(2007)}]{Zenitani_Hoshino-2007}
{Zenitani}, S., \& {Hoshino}, M. 2007, The Astrophysical Journal, 670, 702

\bibitem[{Zweibel \& Yamada(2009)}]{Zweibel}
Zweibel, E.~G., \& Yamada, M. 2009, Annual Review of Astronomy and
  Astrophysics, 47, 291

\end{thebibliography}
\end{document}